\documentclass[a4paper,12pt]{article}
\usepackage{amsmath}
\usepackage{vmargin}
\usepackage{amsthm}

\theoremstyle{plain}
\newtheorem{theorem}{Theorem}
\newtheorem{lemma}{Lemma}

\theoremstyle{remark}
\newtheorem*{remark}{Remark}

\begin{document}

\title{Characteristic invariants and Darboux's method }
\date{}

\author{O.V. Kaptsov and A.V. Zabluda \\
\\
Institute of Computational Modelling SB RAS, \\
Akademgorodok, Krasnoyarsk, 660036, Russia. \\
e-mail: {\tt kaptsov@ksc.krasn.ru} }

\maketitle

\abstract{We develop method that allows to derive reductions and solutions to
hyperbolic systems of partial differential equations. The method is
based on using functions that are constant in the direction of
characteristics of the system. These functions generalize well-known
Riemann invariants. As applications we consider the gas dynamics
system and ideal magnetohydrodynamics equations. In special cases we
find solutions of these equations depending on some arbitrary
functions. }

\section{Introduction}

One of the first methods for finding solutions to nonlinear partial
differential equation
\begin{equation}
 F(x,y,u,u_x,u_y,u_{xx},u_{xy},u_{yy}) = 0          \label{PDE-2-2-nonlin}
\end{equation}
was proposed by Monge and was then further improved by Ampere.
 The method was described in detail in the classical books of
 Goursate \cite{Gour} and  Forsyth \cite{For}.
To apply this method, one must find an equation of the first order
\begin{equation}
 f(x,y,u,u_x,u_y) = c , \qquad c\in R               \label{Monge Method}
\end{equation}
 such that every solution of \eqref{Monge Method} satisfies the equation \eqref{PDE-2-2-nonlin}
  for arbitrary $c$. In this case the function $f$ is called a first integral of the equation
\eqref{PDE-2-2-nonlin}.
 To find first integrals, we
 need to look for functions which are constant in the direction of characteristics of
the equation \eqref{PDE-2-2-nonlin}.
 If there are two first integrals $f_1$ and $f_2$ for given family of
characteristics, then integration of the equation
\eqref{PDE-2-2-nonlin} reduces to solving of the first order
equation
\begin{equation*}
  G(f_1,f_2) = 0 ,
\end{equation*}
with $G$ an arbitrary function.

 In 1870, G. Darboux \cite{Dar} announced a generalization of the
 Monge-Ampere method. He proposed to seek an additional partial differential equation
of second order (or higher)
\begin{equation}
 g(x,y,u,u_x,u_y,u_{xx},u_{xy},u_{yy}) = c         \label{Darboux Method}
\end{equation}
such that the system of equations \eqref{PDE-2-2-nonlin} and
\eqref{Darboux Method} is in involution for all $c$. The function
$g$ turns out to be constant along a family of characteristics of
the equation \eqref{PDE-2-2-nonlin}. In this case, the function $g$
is called a characteristic invariant of \eqref{PDE-2-2-nonlin}. When
partial differential equation \eqref{PDE-2-2-nonlin} has a
sufficient number of characteristic invariants, it can be reduced to
a system of ordinary differential equations. The detailed
description of the Darboux method is also given in the above
mentioned books \cite{Gour} and \cite{For}.

Although equations that are integrable by Darboux's method arise
rarely, they are of much interest. E. Vessiot \cite{Ves1,Ves2}
classified all equations of type
\begin{equation*}
 u_{xy} = w(x,y,u,u_x,u_y)
\end{equation*}
integrable by this method and found a general solution for every
obtained equation. Recently, there was a renewed interest in the
method of Darboux that was studied in \cite{Br}, \cite{And1},
\cite{And2}, \cite{Vas}, \cite{Sok} and \cite{Kap}.

In this paper we consider systems of partial differential equations
in two and $n$ independent variables. In section \ref{Inv.Char} we
introduce an operator of differentiation in the direction of
characteristics of the system and corresponding invariants of
characteristics of order $k$. We prove that if a function $h$
defined on the $k$-th order jet space $J^{(k)}$ is constant  along a
vector field  on the solutions of the system of partial differential
equations, then this function is an invariant of characteristics. In
section \ref{Darboux method} we describe the Darboux method for
systems of partial differential equations in two independent
variables and give its applications to the gas dynamics system and
magnetohydrodynamics equations. We use invariants of characteristics
to reduce these systems and find solutions depending on some
arbitrary functions.

\section{Invariants of characteristics}      \label{Inv.Char}
 Let us begin with a system  of first order partial
differential equations in two independent and $m$ dependent
variables
\begin{equation}
 u_t + F(t,x,u,u_x) = 0 ,                     \label{SPDE-1-2-nonlin-resolved}
\end{equation}
 where  $u=(u^1,\dots,u^m)$, $u_t=(u^1_t,\dots,u^m_t)$,
$u_x=(u^1_x,\dots,u^m_x)$, and $F=(F^1,\dots,F^m)$.

 We denote by $D_t$ and $D_x$ the total derivatives with respect to
 $t$ and $x$. Consider differential operator
\begin{equation}
 D_t + \lambda D_x ,                          \label{Diff.Operator-2-resolved}
\end{equation}
whose coefficient $\lambda$ can depend on $t$, $x$, $u$, and $u_x$.
The operator \eqref{Diff.Operator-2-resolved} is an operator of
differentiation in the direction of characteristics of the system
\eqref{SPDE-1-2-nonlin-resolved}, if the coefficient $\lambda$
satisfies to the equation
\begin{equation}
 \det \Big(\frac{\partial (F)}{\partial (u_x)} -\lambda E \Big) =0 ,        \label{Char.Eqn-resolved}
\end{equation}
where {\Large $\frac{\partial (F)}{\partial (u_x)}=
 \frac{\partial (F^1,\dots,F^m)}{\partial (u^1_x,\dots,u^m_x)}$} is the Jacobi matrix and $E$ is the
identity matrix.

Suppose $ u^i_k$ is the partial derivative of order $k$ of the
function $u^i$ with respect to $x$, then $u_k = (u^1_k,...,u^m_k)$
stands for the vector composed of these derivatives. Let $L$ be an
operator of the differentiation in the direction of characteristics
of the system \eqref{SPDE-1-2-nonlin-resolved}. According to
\cite{Kap}, a function $h(t,x,u,\dots,u_k)$ defined on the $k$-th
order jet space $J^{(k)}$ is called an invariant of characteristics
of order $k$ of the system \eqref{SPDE-1-2-nonlin-resolved}
corresponding to the operator $L$, if $h$ is a solution of the
equation
\begin{equation}
 L(h)|_{[S]} = 0.                                \label{SPDE-Det.Inv.}
\end{equation}
Here [S] means the system \eqref{SPDE-1-2-nonlin-resolved} and its
differential consequences with respect to $x$. When the system
\eqref{SPDE-1-2-nonlin-resolved} has the Riemann invariants, they
are zero order invariants of characteristics.

Some systems have invariants of characteristics of arbitrary order.
For example, consider one-dimensional system of gas dynamics
equations \cite{Cour1}:
\begin{gather}
 u_t + uu_x + p_x/\rho = 0 ,            \nonumber \\
 \rho_t + (\rho u)_x = 0 ,              \label{Gas.1D,nonstat,s-terms} \\
 s_t + us_x = 0 ,                        \nonumber
\end{gather}
where $u$, $\rho$, $p$, and $s$ are the velocity, the density, the
pressure, and the entropy. The equation of state is given by the
function $p = p(\rho, s)$. The operators of the differentiation in
the direction of characteristics of the system
\eqref{Gas.1D,nonstat,s-terms} are
\begin{equation}
 L_1 = D_t + uD_x , \quad
 L_2 = D_t + (u+c)D_x ,\quad            \label{Diff.Operators for Gas.1D,nonstat}
 L_3 = D_t + (u-c)D_x ,
\end{equation}
with $c =$ {\Large $ \sqrt{\frac{\partial p}{\partial \rho} } $} the
speed of sound. Obviously, the entropy $s$ is an invariant of
characteristics corresponding to the operator $L_1$. It is easy to
check that the operator $\displaystyle \frac{1}{\rho}D_x$  commutes
with $L_1$ by virtue of the second equation of the system
\eqref{Gas.1D,nonstat,s-terms}. This implies that the recurrent
formula
\begin{equation*}
 I_{n+1} = \frac{1}{\rho}D_x(I_n) , \quad n=0,1,\dots
\end{equation*}
 gives the invariants of characteristics corresponding
to the operator $L_1$. We will show in section \ref{Darboux method}
that invariants of characteristics corresponding to the operators
$L_2$ and $L_3$ exist only for the special equations of state.

It can be proved \cite{Kap} that if $h_1$ and $h_2$ are invariants
of characteristics of the system \eqref{SPDE-1-2-nonlin-resolved}
corresponding to the operator $L$, then both an arbitrary function
$f(h_1,h_2)$ and $h = $ {\Large $\frac{D_x h_1}{D_x h_2}$} are
invariants of characteristics.

\begin{lemma} Let $L$ be an operator of the form \eqref{Diff.Operator-2-resolved}.
 Suppose that a function $h(t,x,u,u_1,\dots,u_n)$, with $n \ge
 1$, satisfies \eqref{SPDE-Det.Inv.}, then
 $L$ is the operator of the differentiation in
the direction of characteristics of the system
\eqref{SPDE-1-2-nonlin-resolved} and $h$ is an invariant of
characteristics corresponding to the operator $L$.
\end{lemma}
\begin{proof} According to condition of the theorem, $h$ is a
  solution of the equation
\begin{equation}
 D_th +  \lambda D_xh  |_{[S]} = 0 .            \label{SPDE-Det.Inv.,applied}
\end{equation}
 Note that
\begin{equation*}
 D_xh \simeq \sum_{i=1}^{m}u^i_{n+1}h_{u^i_n} ,
\end{equation*}
where the symbol $\simeq$ means that the difference between left and
right-hand sides contains no derivatives of order greater than $n$.
It is easy to see that the formula
\begin{equation*}
\end{equation*}
\begin{equation*}
 D_th \simeq -\sum_{1\leq i,j\leq m}u^j_{n+1}F^i_{u^j_1}h_{u^i_n}
\end{equation*}
is correct because of the system \eqref{SPDE-1-2-nonlin-resolved}.
From the equation \eqref{SPDE-Det.Inv.,applied} we have
\begin{equation*}
 - \sum_{1\leq i,j\leq m}u^j_{n+1}F^i_{u^j_1}h_{u^i_n} +
 \lambda \sum_{i=1}^{m}u^i_{n+1}h_{u^i_n} \simeq 0 .
\end{equation*}
This yields $m$ equations
\begin{equation*}
 \sum_{j=1}^m \Big(F^i_{u^j_1} - \delta^i_j \lambda\Big)h_{u^i_n} =
 0, \qquad  i=1,\dots,m ,
\end{equation*}
with $\delta^i_j$ the Kronecker symbol. Rewriting the above equation
in the matrix form
\begin{equation*}
 \Big(\frac{\partial (F)}{\partial (u_x)} - \lambda E \Big)
 (h_{u^1_n},\dots,h_{u^m_n})^{t} = 0 ,
\end{equation*}
where $(h_{u^1_n},\dots,h_{u^m_n})^{t}$ is the transposed vector, we
conclude that $\lambda$ is a solution of the equation
\eqref{Char.Eqn-resolved}.
\end{proof}

We now consider the system of first order partial differential
equations in $n+1$ independent and $m$ dependent variables
\begin{equation}
 u_t + F(t,x,u,u_{x_1},\dots,u_{x_n}) = 0 ,             \label{SPDE-m-n-nonlin-resolved}
\end{equation}
with $x = (x_1,\dots,x_n)$, $u=(u^1,\dots,u^m)$, $u_{x_i}=
(u^1_{x_i},\dots,u^m_{x_i})$, and $F = (F^1,\dots,F^m)$.

Let us denote by $u_k$ the set of $k$-th order partial derivatives
 of the functions $u^1, \dots, u^m$ with respect to
$x_1, \dots , x_n$. We say that a function \linebreak
$h(t,x,u,u_1,\dots,u_k)$ defined on the $k$-th order jet space
$J^{(k)}$ is an invariant of characteristics of the system
\eqref{SPDE-m-n-nonlin-resolved} if $h$ satisfies to the equation
\begin{equation}
 \det \Big(E D_t h + \sum_{i=1}^m \frac{\partial (F)}{\partial
 (u_{x_i})} D_{x_i}h\Big)|_{[Sn]} = 0 .                 \label{Inv.Char.Eqn-m-n-resolved}
\end{equation}
Here $D_t$ and $D_{x_i}$ are the total derivatives with respect to
$t$ and $x_i$, $[Sn]$ means the system
\eqref{Inv.Char.Eqn-m-n-resolved} and its differential consequences
with respect to $x_i$ ($i=1,\dots,n$), $\displaystyle \frac{\partial
(F)}{\partial (u_{x_i})}= \frac{\partial (F^1,\dots,F^m)}{\partial
(u^1_{x_i},\dots,u^m_{x_i})}$ is the Jacobi matrix and $E$ is an
identity matrix.

An operator
\begin{equation*}
 L = D_t + \lambda_1 D_{x_1} + \dots + \lambda_n D_{x_n} ,
\end{equation*}
where $\lambda_i$ can depend on $t,x,u,u_1$, is called an operator
of differentiation in the direction of the vector field $v
=(1,\lambda_1,\dots,\lambda_n)$.

\begin{theorem} Suppose that there is an operator $L$ of
differentiation in the direction of the vector field $v
=(1,\lambda_1,\dots,\lambda_n)$ and a function
$h(t,x,u,u_1,\dots,u_k)$, with $k\ge 1$, such that
\begin{equation*}
 L(h)|_{[Sn]} = 0  ,
\end{equation*}
then $h$ is an invariant of characteristics of the system
\eqref{SPDE-m-n-nonlin-resolved}.
\end{theorem}

\begin{proof} Since $h$ satisfies
\begin{equation}
 D_t h + \sum_{i=1}^{n}\lambda_iD_{x_i}h |_{[Sn]}= 0 ,       \label{SPDE-Det.Inv.-n}
\end{equation}
the coefficients of $n+1$-th derivatives on the left-hand side of
\eqref{SPDE-Det.Inv.-n} must vanish. To find these coefficients, we
write $D_{x_i}h$ up to $k$-th derivatives:
\begin{equation*}
 D_{x_1}h \simeq \sum_{j=1}^m\sum_{|\alpha|=k} u^j_{\alpha + 1_1}h_{u^j_{\alpha}} ,\quad
 \cdots ,\quad
 D_{x_n}h \simeq \sum_{j=1}^m\sum_{|\alpha|=k} u^j_{\alpha + 1_n}h_{u^j_{\alpha}} .
\end{equation*}
Here $u^j_{\alpha}$ denotes the derivative
 {\Large $\frac{\partial^{|\alpha|} u^j}
 {\partial x^{\alpha_1}_1\cdots\partial x^{\alpha_n}_n }$}
  of order
  $k = |\alpha|= \alpha_1+\cdots +\alpha_n$, $u^j_{\alpha +
 1_i}$ is the derivative  {\Large $\frac{\partial^{|\alpha|+1} u^j}
{\partial x^{\alpha_1}_1\cdots\partial
x^{\alpha_i+1}_i\cdots\partial x^{\alpha_n}_n }$}, and the symbol
$\simeq$ means that the difference between the left- and right-hand
sides contains no the $k+1$-th derivatives. This yields
\begin{equation*}
 \lambda_1D_{x_1}h + \cdots + \lambda_nD_{x_n}h \simeq
 \sum_{i=1}^n\lambda_i \sum_{j=1}^m\sum_{|\alpha|=k} u^j_{\alpha +
 1_i}h_{u^j_{\alpha}} .
\end{equation*}
On the other hand, we have
\begin{equation*}
 D_t h |_{[Sn]} \simeq -\sum_{j=1}^m\sum_{|\alpha|=k}D^{\alpha}(F^j)h_{u^j_{\alpha}}
 \simeq -\sum_{j=1}^m\sum_{|\alpha|=k}
 \Big(\sum_{i=1}^n\sum_{s=1}^m u^s_{\alpha +1_i}F^j_{u^s_{x_i}}\Big)h_{u^j_{\alpha}} .
\end{equation*}
The above calculations lead to
\begin{equation*}
 D_t h + \sum_{i=1}^n\lambda_iD_{x_i}h |_{[Sn]} \simeq
 -\sum_{j=1}^m\sum_{|\alpha|=k} \Big[\sum_{i=1}^n\Big(\sum_{s=1}^m
 u^s_{\alpha +1_i}F^j_{u_{x^s_i}} - \lambda_i u^j_{\alpha +1_i}\Big)\Big]h_{u^j_{\alpha}} = 0 .
\end{equation*}
It is convenient to represent the last relation in a matrix form
\begin{equation}
 \sum_{i=1}^n \sum_{|\alpha| = k} u_{\alpha +1_i}
 A^{x_i} h_{u_{\alpha}}  = 0 ,                      \label{Matrix.Form}
\end{equation}
with $u_{\alpha} = (u^1_{\alpha},\dots,u^m_{\alpha})$,
$h_{u_{\alpha}} = (h_{u^1_{\alpha}},\dots,h_{u^m_{\alpha}})$, and
 $ A^{x_i} = $ {\Large $\frac{\partial (F)}{\partial (u_{x_i})}$  }
$- \lambda_i E$.

We need to prove that $h$ is a solution of the equation
\eqref{Inv.Char.Eqn-m-n-resolved} which is equivalent to the
following:
\begin{equation}
 \det \Big( \sum_{i=1}^n A^{x_i} D_{x_i}h \Big) = 0 .       \label{Det.Matrix.Form}
\end{equation}
 For this purpose, it is enough to show that the linear homogeneous
 system
\begin{equation}
 \Big(\sum_{i=1}^n A^{x_i}D_{x_i}h \Big) r = 0             \label{Lin.Sys}
\end{equation}
has a nontrivial solution $r$. This solution is expressed in the
form
\begin{equation}
 r = \sum_{|\alpha| = k} (Dh)^{\alpha} h_{u_{\alpha}} ,    \label{Lin.Sys.Sol}
\end{equation}
where $(Dh)^{\alpha} = (D_{x_1}h)^{\alpha_1}\cdots
(D_{x_n}h)^{\alpha_n}$.
Indeed, substituting  \eqref{Lin.Sys.Sol} in the left-hand side of
\eqref{SPDE-Det.Inv.-n} leads to
\begin{equation}
 \Big(\sum_{i=1}^n A^{x_i}D_{x_i}h\Big)
 \Big(\sum_{|\alpha| = k}(Dh)^{\alpha}h_{u_{\alpha}}\Big) =          \label{Lin.Sys.Subs}
 \sum_{i=1}^n \sum_{|\alpha| = k} A^{x_i}(Dh)^{\alpha +1_i}h_{u_{\alpha}} .
\end{equation}
Note that, the expressions including $u_{\alpha +1_i}$ in
\eqref{Matrix.Form} coincide with ones including $(Dh)^{\alpha
+1_i}$ in \eqref{Lin.Sys.Subs}. Since the left-hand side of
\eqref{Matrix.Form} is zero then \eqref{Lin.Sys.Subs} is equal to
zero as well. Hence, \eqref{Det.Matrix.Form} is valid.
\end{proof}

When the conditions of the theorem are satisfied, we say that a
function $h$ defined on the $k$-th order jet space $J^{(k)}$ is
constant  along a vector field $v=(1,\lambda_1,\dots,\lambda_n)$ on
the solutions of the system \eqref{SPDE-m-n-nonlin-resolved}.

\section{The Darboux method and its applications} \label{Darboux method}

In this section, we will describe the Darboux method for systems of
partial differential equations and give relevant examples. The
detailed description of applications of this approach to second
order partial differential equations in two independent variables is
given in \cite{Gour}.

The Darboux method is based on using the invariants of
characteristics. Let us consider the system
\eqref{SPDE-1-2-nonlin-resolved} and assume that the corresponding
equation \eqref{Char.Eqn-resolved} has $m$ distinctive real roots
$\lambda_1,\dots,\lambda_m$. If there are two functionally
independent invariants of characteristics  $I_i, J_i$ for all
$\lambda_i$, then we can constitute a system of ordinary
differential equations in the independent variable $x$
\begin{equation}
 f_1(I_1,J_1) = 0 ,\quad \dots,
 \quad f_m(I_m,J_m) = 0,                    \label{Darboux-I-J}
\end{equation}
where $f_1,\dots,f_m$ are arbitrary functions. It is necessary to
look for the general solution of the systems
\eqref{SPDE-1-2-nonlin-resolved} and \eqref{Darboux-I-J}. If we can
find a general solution of \eqref{Darboux-I-J}, then substituting
this solution into \eqref{SPDE-1-2-nonlin-resolved} leads to a
system of ordinary differential equations in the independent
variable $t$. It is enough to solve the last system in order to find
the general solution of \eqref{SPDE-1-2-nonlin-resolved}.

\begin{remark} When the equation \eqref{Char.Eqn-resolved} has
a root of multiplicity $k$ it is desirable to obtain $k+1$
invariants $J_1,\dots,J_{k+1}$ corresponding to this root. In this
case, the system \eqref{Darboux-I-J} includes equations
\begin{equation*}
 f_1(J_2,J_1) = 0 , \dots, f_k(J_{k+1},J_1) = 0 .
\end{equation*}
Such example arises naturally in the equations of
magnetohydrodynamics.
\end{remark}

As example, consider the system of gas dynamics equations in two
independent variables $t$ and $x$:
 \begin{gather}
 u_t + uu_x + p_x/\rho = 0 ,            \nonumber \\
 \rho_t + (\rho u)_x = 0 ,              \label{Gas.1D,nonstat,p-terms}\\
 p_t + up_x + \rho c^2u_x = 0 .         \nonumber
\end{gather}
 where  $\rho$ is density, $u$ is velocity,  $p$ is pressure, and $c(\rho ,p)$  is sound
 speed.

One can easily deduce zero order invariants of characteristics of
the system \eqref{Gas.1D,nonstat,p-terms} corresponding to the
operator $L_2$ (or $L_3$) given by \eqref{Diff.Operators for
Gas.1D,nonstat}. To do so, following \cite{Kap}, one must seek all
solutions of the equation
\begin{equation}
 D_t h + (u + c)D_x h |_{[G]} = 0 ,     \label{SPDE-Det.Inv. (1,u+c)}
\end{equation}
where $[G]$ stands for system \eqref{Gas.1D,nonstat,p-terms} and its
differential consequences with respect to $x$; the function $h$ can
depend on $t, x, u, \rho, p$. Obviously, the left-hand side of
\eqref{SPDE-Det.Inv. (1,u+c)} is a polynomial of the first degree in
$u_x, \rho_x,$  and $p_x$. Collecting similar terms of these
variables leads to the following equations
\begin{equation}
 h_\rho = 0 , \qquad
 h_u = \rho c h_p , \qquad              \label{SPDE-Det.Inv-Gas.1D,nonstat,p-terms}
 h_t + (u+c)h_x = 0 .
\end{equation}
It follows from the first and second equations of
\eqref{SPDE-Det.Inv-Gas.1D,nonstat,p-terms} that a nonconstant
solution $h$ exist only if
\begin{equation}
 c = g(p)/\rho ,                        \label{c-piv-0}
\end{equation}
with $g$ an arbitrary function of $p$. As a consequence of the third
equation, $h$ is independent of $t$ and $x$. According to the second
equation of \eqref{SPDE-Det.Inv-Gas.1D,nonstat,p-terms},  $h$ is an
arbitrary function of
\begin{equation*}
 I^{+} = u + \int\frac{dp}{g(p)} .
\end{equation*}
 Similarly, it is possible to check that the Riemann invariant
\begin{equation*}
 I^{-} = u - \int\frac{dp}{g(p)}
\end{equation*}
corresponds to the operator $L_3$.

We now use the invariant $I^{-}$ to derive solutions of the system
\eqref{Gas.1D,nonstat,p-terms}. Setting $I^{-} = 0$ and introducing
a new function $\displaystyle F= \int \frac{dp}{g(p)}$, we get the
following representation
\begin{equation*}
 u = F(p) .
\end{equation*}
 In this case, the system \eqref{Gas.1D,nonstat,p-terms} reduces to
\begin{equation}
 \rho_t + \left(F\rho\right)_x = 0, \qquad
 p_t + Fp_x + \frac{p_x}{F'\rho} = 0.           \label{Gas.1D,nonstat,p-terms,reduced}
\end{equation}
The previous system admits the Riemann invariants
\begin{equation*}
 p, \qquad r = \int\left(F'\right)^2 dp + \frac{1}{\rho} .
\end{equation*}
One then can rewrite the system
\eqref{Gas.1D,nonstat,p-terms,reduced} in the form
\begin{equation*}
 p_t + \left(F(p)-\frac{G(p)-r}{F'(p)}\right)p_x = 0, \qquad
 r_t + F(p)r_x= 0,
\end{equation*}
where $\displaystyle G(p) = \int{\big(F'(p)\big)^2 dp}$. Using the
hodograph transformation leads to the linear system
\begin{equation*}
 x_p - F(p)t_p = 0, \qquad
 \big(F(p)F'(p)-G(p)+r\big)t_r - F'(p)x_r = 0.
\end{equation*}
The general solution of this system  is
\begin{equation}
 t = P + R'F' , \qquad                              \label{Gas.1D,nonstat,p-terms,reduced,Sln}
 x = R'\left(FF'+1/\rho\right) - R + \int FP'dp ,
\end{equation}
where $P=P(p)$ and  $R=R(r)$ are arbitrary functions. As a
consequence of \eqref{Gas.1D,nonstat,p-terms,reduced,Sln}, we obtain
a solution of \eqref{Gas.1D,nonstat,p-terms} in the implicit form.

It was shown in \cite{Kap} that the first order invariants
corresponding to the operators $L_2$ and $L_3$ exist only if the
speed sound is given by
\begin{equation*}
 c=(a+bp)^{(2/3)}/\rho , \qquad a,b\in R.
\end{equation*}
The corresponding equation of state is
\begin{equation*}
 p(\rho,s) = -\frac{1}{b}
  \left[ a + \left(\frac{3\rho}{b\left(A(s)\rho-1\right)}\right)^3 \right] ,
\end{equation*}
the Riemann invariants are
\begin{equation*}
 I_2  = bu+3(a+bp)^{1/3} , \qquad
 I_3 = bu-3(a+bp)^{1/3} ,
\end{equation*}
 and the first order invariants have the form
\begin{equation*}
 J_2 = \frac{\rho(a+bp)^{1/3}}{u_x(a+bp)^{2/3} + p_x} - \frac{bt}{3} , \qquad
 J_3 = \frac{\rho(a+bp)^{1/3}}{u_x(a+bp)^{2/3} - p_x} - \frac{bt}{3} .
\end{equation*}
The invariants corresponding to the operator $L_1 = D_t + uD_x$
(mentioned in the section \ref{Inv.Char}) are
\begin{equation*}
 I_1  = b/\rho-3(a+bp)^{-1/3} , \qquad
 J_1 = 1/\rho D_x(I_1).
\end{equation*}
The gas dynamics system is conveniently written in terms of the
Riemann invariants
\begin{equation}
 \begin{split}                                         \label{Gas.1D,Riem}
  (I_1)_t & = -\frac{I_2 + I_3}{2b} (I_1)_x, \\
  (I_2)_t & = \frac{I_1(I_2^2 - I_3^2) - 2 I_2 M}{36b} (I_2)_x, \\
  (I_3)_t & = \frac{I_1(I_2^2 - I_3^2) - 2 I_3 M}{36b} (I_3)_x,
 \end{split}
\end{equation}
 with $M = I_1(I_2-I_3)+18$. The first order invariants $J_1 ,
 J_2$, and $J_3$ can be represented as
\begin{equation}
 J_1  = \frac{M}{I_2-I_3}(I_1)_x , \qquad
 J_2  = t-\frac{18b}{M (I_2)_x} , \qquad        \label{J(I)}
 J_3  = t-\frac{18b}{M (I_3)_x} .
\end{equation}

We now apply the Darboux method to reduce the system
\eqref{Gas.1D,Riem} to some ordinary differential equations. The
corresponding system \eqref{Darboux-I-J} is equivalent to
\begin{equation}
 J_1 =  F_1(I_1) , \qquad
 J_2 =  F_2(I_2) , \qquad                        \label{J=F(I)}
 J_3 =  F_3(I_3) ,
\end{equation}
where $F_1$, $F_1$, and $F_3$ are arbitrary functions. From
\eqref{Gas.1D,Riem}, \eqref{J(I)} and \eqref{J=F(I)} we get two
systems of ordinary differential equations:
\begin{align}
 (I_1)_x & = \frac{F_1(I_1)(I_2 - I_3)}{M} ,            \nonumber \\
 (I_2)_x & = \frac{18b}{M\left(t-F_2(I_2)\right)} ,     \label{ODE_x}\\
 (I_3)_x & = \frac{18b}{M\left(t-F_3(I_3)\right)},      \nonumber
\end{align}
\begin{align}
 (I_1)_t & = -\frac{F_1(I_1)(I_2^2-I_3^2)}{2Mb},                   \nonumber \\
 (I_2)_t & = \frac{I_1(I_2^2 - I_3^2) - 2 I_2 M}{2M(t-F_2(I_2))},  \label{ODE_t}\\
 (I_3)_t & = \frac{I_1(I_2^2 - I_3^2) - 2 I_3 M}{2M(t-F_3(I_3))} . \nonumber
\end{align}

Introducing new functions
\begin{equation*}
 \Psi(I_1) = \int{\frac{bI_1}{F_1(I_1)} dI_1},  \qquad
 G_i(I_i) = \int{F_i(I_i) dI_i}, \qquad i=2,3,
\end{equation*}
one may write the equations \eqref{ODE_x} in the following way
\begin{equation*}
 \left[bx-\Psi(I_1)\right]_x = \frac{18b}{M}, \qquad
 \left[tI_2-G_2(I_2)\right]_x = \frac{18b}{M},\qquad
 \left[tI_3-G_3(I_3)\right]_x = \frac{18b}{M} .
\end{equation*}
 Hence, the system \eqref{ODE_x} has the first integrals
\begin{align*}
 tI_2-G_2(I_2)-bx+\Psi(I_1) & = c_2(t), \\
 tI_3-G_3(I_3)-bx+\Psi(I_1) & = c_3(t),
\end{align*}
with $c_2(t)$ and $c_3(t)$ arbitrary functions. Differentiating the
previous relations with respect to $t$ and using the system
\eqref{ODE_t}, we deduce that the functions $c_2(t)$ and $c_3(t)$
are constants.

Therefore, the system \eqref{Gas.1D,Riem} can be reduced to a couple
of differential equations
\begin{equation}
\begin{split}       \label{Gas.1D,Sln}
 (I_1)_x &= \frac{bI_1(I_2 - I_3)}{\Psi'(I_1)\left(I_1(I_2-I_3)+18\right)}, \\
 (I_1)_t &= -\frac{I_1(I_2^2 - I_3^2)}{2\Psi'(I_1)\left(I_1(I_2-I_3)+18\right)},
\end{split}
\end{equation}
 where $I_2$ and $I_3$ must be expressed from relations
\begin{equation*}
 tI_2-G_2(I_2)-bx+\Psi(I_1)  = 0, \qquad
 tI_3-G_3(I_3)-bx+\Psi(I_1)  = 0 .
\end{equation*}
 It is possible to find solutions of gas dynamics equations by integrating the equations \eqref{Gas.1D,Sln}
with partial functions $G_2, G_3$, and $\Psi$.

It is interesting to note that there are the second order invariants
of characteristics
\begin{gather*}
 I^{\pm}_{(4/5)} = \frac{3}{5}t +
        \frac{p^{1/5}\left(5\rho p p_{xx} \pm 5p^{9/5}u_xx - 5p\rho_x p_x \mp 5p^{9/5} - p^{8/5}\rho u_x^2 - 3\rho p_x^2 \right)}
             {\left(p^{4/5}u_x \pm p_x\right)^3}, \\
 I^{\pm}_{(2)} = \frac{Gp^3(\rho u_x G \pm G_x)}{\rho},
\end{gather*}
with $\displaystyle G=\frac{\rho}{p^2 u_x \pm p_x}$, corresponding
to the operators $L^{\pm} = D_t + (u \pm c)D_x$, when the speed
sound is given by one of the following formulas
\begin{equation*}
 c = p^{4/5}/\rho , \qquad
 c = p^2/\rho .
\end{equation*}

We now consider the one-dimensional magnetohydrodynamics equations
\cite{Pai}
\begin{gather}
 \rho_t + (\rho u)_x = 0 ,          \nonumber\\
 \displaystyle u_t + uu_x + \frac{p_x}{\rho} + \frac{({B_2}^2+{B_3}^2)_x}{4 \pi \rho}= 0 ,  \nonumber\\
 \displaystyle v_t + uv_x = 0 ,     \label{MGD.1D,nonstat}\\
 \displaystyle w_t + uw_x = 0 ,     \nonumber\\
 (B_2)_t + (u B_2)_x = 0,           \nonumber\\
 (B_3)_t + (u B_3)_x = 0,           \nonumber\\
 s_t + us_x =0.                     \nonumber
\end{gather}
 Here $\rho$ is the density, $p$ is the pressure, and $s$ is the entropy; $(u,v,w)$ and $(B_1,B_2,B_3)$ denote
the velocity and the magnetic fields, respectively. We assume that
$B_1 = 0$ and $p$ is a function of $\rho$ and $s$. In this case,
there are the following invariants of characteristics
\begin{equation*}
  I_1 = s, \quad
  I_2 = v, \quad
  I_3 = w, \quad
  I_4 = \frac{B_2}{\rho}, \quad
  I_5 = \frac{B_3}{\rho} ,\quad
  J   = \frac{s_x}{\rho} ,
\end{equation*}
corresponding to the operator $L_1 =D_t+uD_x$.

Using these invariants we will reduce the system
\eqref{MGD.1D,nonstat} to the second order equation for the entropy.
According to the general scheme of the Darboux method, we write
\begin{equation*}
 v = V(s), \quad
 w = W(s), \quad
 s_x/\rho = f_1(s), \quad
 B_2/\rho = f_2(s), \quad
 B_3/\rho = f_3(s),
\end{equation*}
where $V, W, f_1, f_2$, and $f_3$ are arbitrary functions. These
relations are equivalent to the following representation
\begin{equation}
 v = V(s), \quad
 w = W(s), \quad
 \rho = s_x \Pi(s), \quad        \label{MGD,Rep}
 B_2 = s_x \Phi(s), \quad
 B_3 = s_x F(s) ,
\end{equation}
 with $\Pi(s) = 1/f_1(s)$, $\Phi(s) = f_2(s)/f_1(s)$, and $F(s) = f_3(s)/f_1(s)$.

From the last equation of \eqref{MGD.1D,nonstat} we get
\begin{equation}
  u = -\frac{s_t}{s_x} .         \label{MGD.1D,nonstat(last)}
\end{equation}
Substituting \eqref{MGD,Rep} and \eqref{MGD.1D,nonstat(last)} into
the system \eqref{MGD.1D,nonstat} one can check that six equations
of \eqref{MGD.1D,nonstat} are satisfied identically and only the
second equation leads to
\begin{multline}
  -4 \pi {s_x}^2 \Pi(s) s_{tt} + 8 \pi s_t s_x \Pi(s) s_{tx} \\
  + \left(4 \pi {s_x}^2 \Pi(s) p^{\prime}_{\rho} + {s_x}^3 {F(s)}^2 -
  4 \pi {s_t}^2 \Pi(s) + {s_x}^3 {\Phi(s)}^2 \right)s_{xx}\\             \label{MGD,nonstat(second)}
   + 4 \pi {s_x}^3 p^{\prime}_{s} + 4 \pi {s_x}^4 p^{\prime}_{\rho} \Pi'(s) +
    {s_x}^5 \Phi(s) \Phi'(s) + {s_x}^5 F(s) F'(s)=0 .
\end{multline}
Suppose that the pressure has the following form
\begin{equation*}
 p = G(s) - \frac{{\Phi(s)}^2+{F(s)}^2}{8 \pi {\Pi(s)}^2} \rho^2,
\end{equation*}
where $G$ is an arbitrary function, then the equation
\eqref{MGD,nonstat(second)} reduces to
\begin{equation}
 {s_x}^2 s_{tt} - 2 s_t s_x s_{tx} + {s_t}^2 s_{xx} -
 \frac{G'(s)}{\Pi(s)}s_x^3= 0 .                                         \label{MGD,nonstat(second),Reduced}
\end{equation}
It is possible to find two intermediate integrals \cite{Gour} for
the equation \eqref{MGD,nonstat(second),Reduced}
\begin{equation}
 \frac{s_t}{s_x}-\frac{G'(s)t}{\Pi(s)} = \phi(s) ,\qquad
 x - \frac{s_t}{s_x}t + \frac{G'(s)}{2\Pi(s)}t^2 = \psi(s),             \label{Interm.Integrals}
\end{equation}
 with $\phi$ and $\psi$ arbitrary functions. Eliminating $s_t/s_x$
from \eqref{Interm.Integrals} gives the implicit solution of
\eqref{MGD,nonstat(second),Reduced}
\begin{equation}
 x + \phi(s)t + \frac{G'(s)}{2\Pi(s)}t^2 = \psi(s).                     \label{MGD,nonstat(second),Sln}
\end{equation}
In the special cases one can express $s$ from
\eqref{MGD,nonstat(second),Sln} and find the explicit solutions of
the one-dimensional magnetohydrodynamics equations
\eqref{MGD.1D,nonstat}.

\section{Conclusion}
In this article we have developed the method of integrating system
of first order partial differential equations in two independent
variables. This method can be extended to the hyperbolic system of
high order equations. It is important that corresponding invariants
should exist for every family of characteristics of the system.

When the system includes equations in $n$ ($n\ge 3$) independent
variables then we face a difficult task. As usual these systems have
few invariants of characteristics. For example, the two-dimensional
unsteady gas dynamics equations admit only one invariant of
characteristics, namely the entropy. In three-dimensional case, the
Ertel's integral is an additional invariant of characteristics. Note
that in the case of the two-dimensional steady gas dynamics
equations, Kaptsov \cite{Kap} have founded a first order invariant
\begin{equation*}
 J_0 = \frac{D_y(I_B)}{s_y},
\end{equation*}
where $I_B$ is Bernoulli's integral
\begin{equation*}
 I_B = \frac{u^2 + v^2}{2} + \int \frac{1}{\rho} p^{\prime}_{\rho}
  d\rho .
\end{equation*}
 Here, as usual, $u$ and $v$ are the components of velocity, $p$ is
the pressure, $\rho$ is the density, and $s$ is the entropy.

There are problems that we have not yet studied. For example, it is
interesting to look for new solutions of the one-dimensional gas
dynamics equations using the second order invariants
($I^{\pm}_{(4/5)}$ and $I^{\pm}_{(2)}$), give interpretation to some
founded solutions, and apply the Darboux method to other hyperbolic
systems. To simplify calculations, we have implemented the package
of analytical computations which calculates characteristics and
their invariants for the given system.


\section*{Acknowledgments}
The work is supported by the Russian Foundation for Basic Research
(No. 04--01--00130).

\end{document}